\documentclass[twocolumn,preprintnumbers,amsmath,amssymb,superscriptaddress]{revtex4}

\usepackage{bm}%

\usepackage[dvips]{graphicx, color}
\usepackage{subfigure}
\usepackage{comment}

\expandafter\ifx\csname package@font\endcsname\relax\else
 \expandafter\expandafter
 \expandafter\usepackage
 \expandafter\expandafter
 \expandafter{\csname package@font\endcsname}%
\fi

\begin{document}

\title{Enhanced ferromagnetic moment in Co-doped BiFeO$_3$ thin films studied by soft X-ray circular dichroism }%

\author{V.~R.~Singh}
\thanks{\it Presently: Department of Physics and Astronomy, University of Nebraska-Lincoln, NE 68588-0299, USA}
\affiliation{Department of Physics, University of Tokyo,Bunkyo-ku, Tokyo 113-0033, Japan}

\author{V.~K.~Verma}
\affiliation{Department of Physics, University of Tokyo,Bunkyo-ku, Tokyo 113-0033, Japan}

\author{K.~Ishigami}
\affiliation{Department of Physics, University of Tokyo,Bunkyo-ku, Tokyo 113-0033, Japan}

\author{G.~Shibata}
\affiliation{Department of Physics, University of Tokyo,Bunkyo-ku, Tokyo 113-0033, Japan}

\author{Y.~Yamazaki}
\affiliation{Department of Physics, University of Tokyo,Bunkyo-ku, Tokyo 113-0033, Japan}

\author{A.~Fujimori}
\email{fujimori@phys.s.u-tokyo.ac.jp}
\affiliation{Department of Physics, University of Tokyo,Bunkyo-ku, Tokyo 113-0033, Japan}
\affiliation{Condensed Matter Science Division, JAEA, Sayo-cho, Sayo-gun, Hyogo, 679-5148, Japan}

\author{Y.~Takeda}
\affiliation{Condensed Matter Science Division, JAEA, Sayo-cho, Sayo-gun, Hyogo, 679-5148, Japan}

\author{T.~Okane}
\affiliation{Condensed Matter Science Division, JAEA, Sayo-cho, Sayo-gun, Hyogo, 679-5148, Japan}

\author{Y.~Saitoh}
\affiliation{Condensed Matter Science Division, JAEA, Sayo-cho, Sayo-gun, Hyogo, 679-5148, Japan}

\author{H.~Yamagami}
\affiliation{Condensed Matter Science Division, JAEA, Sayo-cho, Sayo-gun, Hyogo, 679-5148, Japan}

\author{Y.~Nakamura}
\affiliation{Institutes for Chemical Research, Kyoto University, Gokasho, Uji, Kyoto 611-0011, Japan}
\author{M.~Azuma }
\affiliation{Institutes for Chemical Research, Kyoto University, Gokasho, Uji, Kyoto 611-0011, Japan}
\author{Y.~Shimakawa}
\affiliation{Institutes for Chemical Research, Kyoto University, Gokasho, Uji, Kyoto 611-0011, Japan}
\date{\today}

\begin{abstract}

 \ \ \ BiFeO$_3$ (BFO) shows both ferroelectricity and magnetic ordering at room temperature but its ferromagnetic component, which is due to spin canting, is negligible. Substitution of transition-metal atoms such as Co for Fe is known to enhance the ferromagnetic component in BFO. In order to reveal the origin of such magnetization enhancement, we performed  soft x-ray absorption spectroscopy (XAS) and soft x-ray magnetic circular dichroism (XMCD) studies of BiFe$_{1-x}$Co$_x$O$_3$ ({\it x} = 0 to 0.30) (BFCO) thin films grown on LaAlO$_3$(001) substrates. The XAS results indicated that the Fe and Co ions are in the Fe$^{3+}$ and Co$^{3+}$ states. The XMCD results showed that the Fe ions show  ferromagnetism while the Co ions are antiferromagnetic at room temperature. The XAS and XMCD measurements also revealed that part of the Fe$^{3+}$ ions are tetrahedrally co-ordinated by oxygen ions but that the XMCD signals of the octahedrally coordinated Fe$^{3+}$ ions increase with Co content. The results suggest that an impurity phase such as the ferrimagnetic $\gamma$-Fe$_2$O$_3$ which exists at low Co concentration decreases with increasing Co concentration and that the ferromagnetic component of the Fe$^{3+}$ ion in the octrahedral crystal fields increases with Co concentration, probably reflecting the increased canting of the Fe$^{3+}$ ions.

\end{abstract}


\maketitle
\section{Introduction}

\ \ \ Multiferroics, which simultaneously show spontaneous electric and magnetic ordering in a single phase \cite{1,2,3,4}, have attracted tremendous interest in recent years because of their potential applications in the field of spintronics such as information storage and sensors, and also because of their underlying fascinating fundamental physics \cite{1}. There are not many multiferroic materials in nature at room temperature (RT), however, because of the rarity of the co-existence of the conventional cation off-center distortion mechanism of ferroelectrics and the formation of magnetic moments at the cation sites \cite{2}.  BiFeO$_3$ (BFO) is such a rare case. However, the magnetic order in the cycloidal antiferromagnetic structure with a only tiny ferromagnetic component, due to spin canting, leads only to a weak magnetoelectric coupling \cite{7,8,34B,34C}.

\ \ \ In the case of thin films, it has been shown that heteroepitaxially strained BFO films are ferromagnetic at RT and show remarkably larger magnetoelectric effect \cite{9}. Bai {\it et al.} \cite{10} reported multiferroic behavior of strained BFO thin films with both large ferromagnetic and ferroelectric polarizations at RT. This suggests the potential of BFO thin films for device applications that exploit the large magnetoelectric coupling. Recently, enhancement of the magnetization was reported for BiFe$_{1-x}$Co$_x$O$_3$ (BFCO) thin films \cite{11,12}. Some of the present authors have systematically studied the structural transition on bulk BFCO samples as well as the ferroelectric and piezo electric properties of thin films fabricated by a chemical solution deposition (CSD) method \cite{13, 14,15}.

\ \ \ In order to investigate the mechanism of the magnetization enhancement by Co doping, X-ray absorption spectroscopy (XAS) and magnetic circular dichroism (XMCD) are powerful tools because they are element specific probes of the electronic and magnetic properties applicable to complex materials.  Several studies have utilized these techniques to examine the electronic and magnetic states of Fe in BFO and its heterostructures \cite{16,17,18,19}. However, measurements of BFO films by Singh {\it et al.} \cite{20} and Bea {\it et al.} \cite{21} showed that the observed magnetic moment might be due to the formation of impurity phases such as the ferrimagnetic $\gamma$-Fe$_2$O$_3$.

\ \ \ The structure and properties of bulk single crystals of BFO have been extensively studied \cite{13, 16, 17, 18, 19, 22}. The crystal have a rhombohedrally distorted perovskite structure (a = b = c = 5.63 \AA, $\alpha$ = $\beta$ = $\gamma$ = 59.4$^\circ$) at RT. The distorted perovskite-type BFO is both ferroelectric ($T_C$ $\sim$1103 K) and antiferromagnetic ($T_N$ $\sim$ 643 K) at RT, and exhibits weak ferromagnetism because of the canted spin structure \cite{23}. The magnetic structure is nearly $G$-type, i.e., the Fe magnetic moments are coupled ferromagnetically within the pseudocubic (111) plane and antiferromagnetically between adjacent planes with an additional incommensurate spin modulation prohibiting the linear magnetoelectric effect from being observed \cite{24}.  It has also been predicted that spontaneous magnetization can be induced in BFO by mixed valence or chemical substitution either changing the Fe-O-Fe bond angle or a statistical distortion of the FeO$_6$ octahedra\cite{1,4,10,23,24,25,26,27}.

\ \ \ BiCoO$_3$ (BCO) also exhibits high transition temperatures above RT with respect to both magnetism and ferroelectricity.  BCO exists as a high pressure phase and can be synthesized under a high pressure of 6 GPa \cite{28}. It has a tetragonal structure with a large c/a ratio of 1.29. The spontaneous polarization is calculated to be 150 $\mu$C/cm$^{-2}$  but no transition to paraelectric phase is observed below the decomposition temperature, and the Neel temperature is 473 K \cite{28}.

\ \ \ In the present paper, we have performed XAS and XMCD studies of the composition dependence of the electronic and magnetic properties of BFCO thin films grown on single-crystal LaAlO$_3$ (LAO) (001) substrates using the CSD method.  Based on the experimental results, we shall discuss the origin of ferromagnetism in the BFCO thin films.

\section{Experimental}

\ \ \ BFCO thin films were grown on LAO (001) substrates using the CSD technique.  The substrates were spin-coated (at 4000 rpm for 30 s) using a solution containing Bi, Co, and Fe organometallic compounds (Koujundo Chemical Laboratory Co., Ltd.) in a xylene carrier in the desired ratio at a concentration of 0.2 mol/kg. The samples were then heated in air at 250$^\circ$C for 5 min. Repeating this coating-heating sequence 13 times yielded BFCO films that were about 260 nm thick. After the films were annealed for 60 min in a 5\% ozone-oxygen mixture in a tube furnace at 420$^\circ$C, the BFCO/LAO(001) samples were subject to again annealing at 435$^\circ$C for 30 min to improve their crystallinity. The stoichiometry was confirmed by HRTEM$-$EDX and was almost constant within the deviation of $\pm$1.0\% \cite{13}.

\ \ \ Structural characterization was carried out using x-ray diffraction (XRD), which demonstrated a clear single phase \cite{13,14}. The XRD measurements confirmed that, as the Co content increases, BFCO turns from rhombohedral (R)  to tetragonal-like monoclinic crystal structures (T) and both phases (R and T) coexist for Co content between $x$ = 0.10 and 0.20.

\ \ \ The XAS and XMCD measurements were performed at the JAEA undulator beam line BL-23SU of SPring-8. The measurements were performed at RT in an applied magnetic field up to $\pm$ 3T. The propagation vector of light with circular polarization was parallel to the magnetic field and incident perpendicular to the sample surface. The spectra were collected in the total electron yield (TEY) mode, which has a  probing depth of $\sim$5 nm. The monochromator resolution was \textsl {E}/$\Delta$\textsl{E}$>$10000, and the degree of circular polarization of the x-rays was 95\% $\pm$ 4\%. The base pressure of the spectrometer chamber was about $10^{-9}$ Torr.

\section{Results and discussion}

\begin{figure}[htbp]
\begin{center}
\includegraphics[width=08cm]{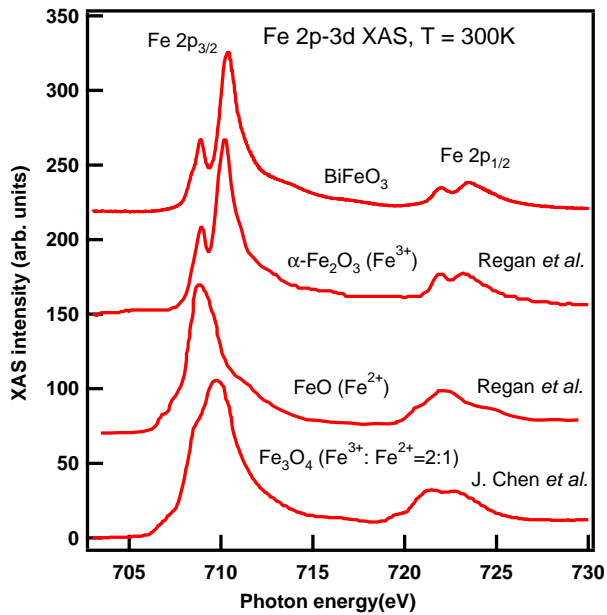}
\caption{(Color online) Fe 2$p$ XAS spectra of BiFeO$_3$ thin film compared with those of $\alpha$-Fe$_2$O$_3$ \cite{29}, FeO \cite{29}, and Fe$_3$O$_4$ \cite{30}.}
\end{center}
\end{figure}
\ \ \ Figure 1 shows the Fe 2$p$ XAS spectra of a BFO compared with those of $\alpha$-Fe$_2$O$_3$ (Fe$^{3+}$) thin film\cite{29},FeO (Fe$^{2+}$) \cite {29}, and Fe$_3$O$_4$ (Fe$^{3+}$:Fe$^{2+}$ = 2:1) \cite {30}. The 2$p_{3/2}$ ($L_3$) edge of the XAS spectrum of BFO  show a two-peak structure  similar to $\alpha$-Fe$_2$O$_3$, indicating that the Fe atoms in BFO are in the trivalent state \cite{18,31}.  The present experimental results (including the XMCD results described below) do not indicate the presence of Fe$^{2+}$ ions, contrary to the previous XAS and XMCD studies of BFO thin films \cite{17}.

 \begin{figure}[htbp]
\begin{center}
\includegraphics[width=08cm]{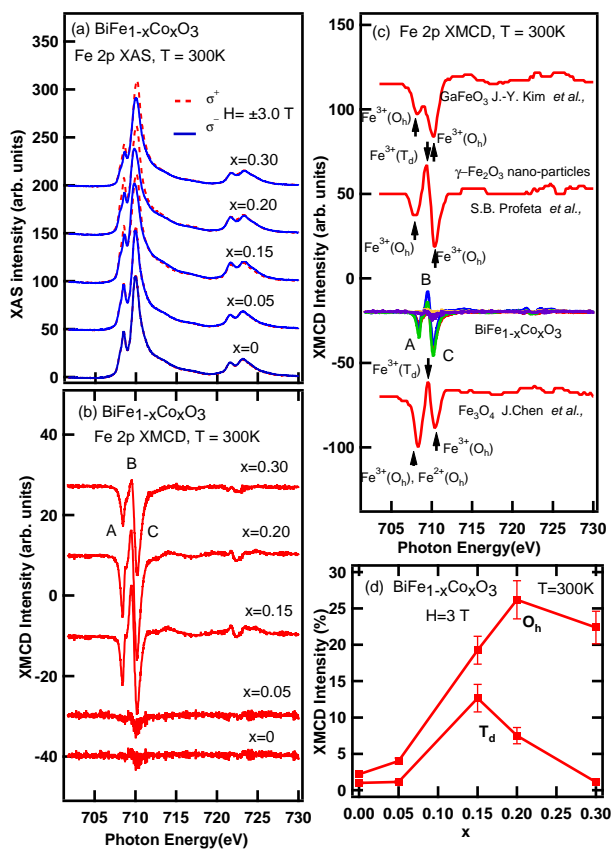}

\caption{(Color online) Fe 2$p$  XAS and XMCD spectra of BiFe$_{1-x}$Co$_x$O$_3$ thin films. (a) XAS spectra of BiFe$_{1-x}$Co$_x$O$_3$ ($x$ = 0 to 0.30) thin films in magnetic fields of $\pm$ 3T at 300 K. (b) XMCD spectra of the BiFe$_{1-x}$Co$_x$O$_3$ ($x$ = 0 to 0.30) thin film in magnetic fields of $\pm$ 3T at 300 K. (c) Comparison of the experimental XMCD spectra of BFCO with those of GaFeO$_3$ \cite{33}, $\gamma$-Fe$_2$O$_3$ \cite{34}, Fe$_3$O$_4$\cite{30}, and FeO \cite{30}. (d) Composition dependence of the heights of peak C due to Fe$^{3+}$ ($O_h$) and peak B due to Fe$^{3+}$ ($T_d$)in the XMCD spectra.}
\end{center}
\end{figure}

\ \ \ Figures 2(a) shows the Fe 2$p$ XAS spectra of the BFCO thin film samples recorded using circularly polarized x-rays. Their difference spectra, i.e., XMCD spectra, are plotted in Figs. 2(b). Here, the XAS spectra obtained with applied magnetic fields of +3.0 and -3.0 T are denoted by $\sigma_+$ and $\sigma_-$, respectively. Sharp negative, positive and negative peaks (denoted by A, B and C) occur in the 2$p_{3/2}$ edge of the XMCD spectra, respectively, around $h\nu$  = 708.5, 709.7, and 710.5 eV. In Fig. 2(c), the XMCD spectrum of BFCO is compared with those of other Fe compounds. The figure shows that the Fe 2$p$ XMCD spectrum of BFCO films were different from those of Fe metal \cite{32}, indicating that the magnetism of the present samples did not arise from Fe metal segregation. Also,  the XMCD spectra of BFCO films are compared with those of GaFeO$_3$ (GFO), where the Fe$^{3+}$ ions are located at the distorted octahedral ($O_h$) sites \cite{33}, $\gamma$-Fe$_2$O$_3$ nanoparticles, where the Fe$^{3+}$ ions are located both at the $O_h$ and tetrahedral ($T_d$) sites with the ratio 5:3 \cite{34}, and Fe$_3$O$_4$, where Fe$^{3+}$ ions are located  both at the $T_d$ and $O_h$ sites (with the ratio 1:1) and Fe$^{2+}$ ions are also located at the $O_h$ sites \cite{30}. The XMCD spectrum of Fe$_3$O$_4$ and that of GFO with Fe$^{3+}$ ions at the $O_h$ sites are clearly different from the spectra of BFCO thin films. On the other hand, the spectral line shape of $\gamma$-Fe$_2$O$_3$, where the XMCD signals arise from the antiferromagnetically coupled  Fe$^{3+}$ ($O_h$) and Fe$^{3+}$($T_d$), is similar to that of the BFCO thin films for samples with small $x$ ($<$ 0.15).

 \ \ \ However, Fig. 2(b) shows that the XMCD spectra change their line shapes with Co content. The negative peaks due to Fe$^{3+}$ at the $O_h$ site increases more rapidly than the positive peak due to Fe$^{3+}$ at the $T_d$ site. In order to see how the XMCD line shape changes with composition, we have plotted the height of the negative peak C [due to Fe$^{3+} $($O_h$)] and the positive peak B [due to Fe$^{3+}$($T_d$)] in Fig. 2(d). The plot indicates that, for low Co content ($x \leq$ 0.15), the height of the $T_d$ peak is as large as that of the $O_h$ peak, while it decreases above $x$ = 0.20. We have therefore decomposed the measured Fe 2$p$ XAS and XMCD spectra into the spectrum of the $\gamma$-Fe$_2$O$_3$ -like impurity phase and that of the Fe$^{3+}$ ion in the O$_h$ crystal field of the BFCO lattice, as plotted in Fig.3(c).  (Details of the decomposition procedure is described in Appendix.). The plot indicates that the $\gamma$-Fe$_2$O$_3$-like phase is dominant for $x <$ 0.15  but starts to decrease for $x >$ 0.15  and the Fe$^{3+}$ ion at the octahedral site becomes dominant for $x >$ 0.20 probably because the canting angle of the Fe$^{3+}$ spins is increased with $x$. Our TEM studies of the BFCO films have suggested that for $x >$ 0.15 the tetragonal-like monoclinic phase exists only near the substrate and that the surface region of these films are dominated by the randomly oriented rhombohedral phase \cite{13}. The rhombohedral phase is responsible for the ferromagnetism through probably the increase of the canting angle of the Fe$^{3+}$ spins with $x$. However, the tetragonal-like monoclinic phase does not contribute to the magnetization \cite{34B,34C}. Therefore, we conclude that the increased ferromagnetism of Fe$^{3+}$ of BFCO is mainly because of the increased canting angle of the Fe$^{3+}$  spins within the rhombohedral phase.

\begin{figure}[htbp]
\begin{center}
\includegraphics[width=05cm]{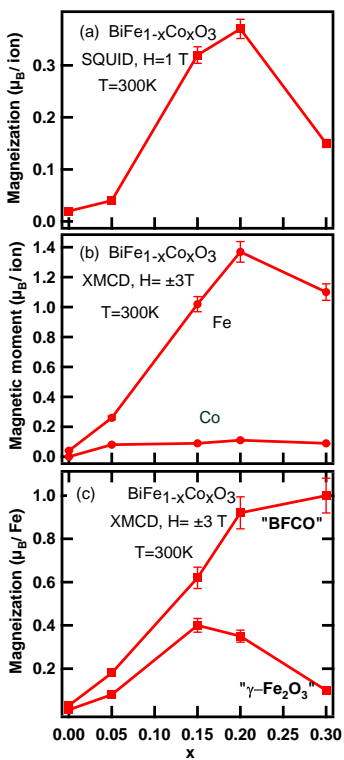}
\caption{(Color online) (a) Magnetization of BiFe$_{1-x}$Co$_x$O$_3$ ($x$ = 0 to 0.30) thin films as a function of Co content in BiFeO$_3$ obtained using a SQUID. (b) Magnetization of Fe and Co in BiFe$_{1-x}$Co$_x$O$_3$ ($x$ = 0 to 0.30) thin films as a function of Co content obtained from XMCD. (c) Decomposition of the Fe magnetic moment using the Fe 2$p$ XAS and XMCD data into the $\gamma$-Fe$_2$O$_3$-like impurity and the Fe ions in the BFCO lattice (See Appendix).}
\end{center}
\end{figure}

\begin{figure}[htbp]
\begin{center}
\includegraphics[width=05cm]{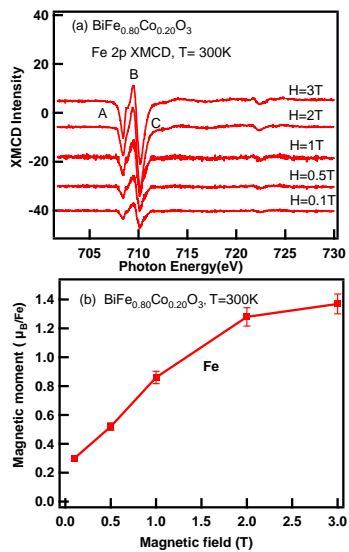}
\caption{(Color online) Magnetic field dependence of the Fe 2$p$ XMCD spectra of the BiFe$_{0.80}$Co$_{0.20}$O$_3$ thin film. (a) Spectra measured at various magnetic fields. (b) Magnetic moment of Fe obtained from the XMCD intensity as a function of magnetic field.}
\end{center}
\end{figure}

\ \ \ The total magnetic moment of Fe in the  BiFe$_{1-x}$Co$_{x}$O$_{3}$ thin films at H=3 T obtained from the XMCD sum rule is plotted as a function of $x$ in Fig. 3(a) and the magnetization at H=1T measured using a SQUID magnetometer in Fig. 3(b). Because the magnetization is strongly magnetic field dependent [Fig.4(b)] and the very different magnetic fields were used for the XMCD (H=3 T) and SQUID (H=1 T) measurements, the magnitude of the the magnetic moment is larger for XMCD than for SQUID. The XMCD data and the magnetization show nearly the same trend. Figure 4(a) shows the Fe 2$p$ XMCD spectrum of BiFe$_{0.80}$Co$_{0.20}$O$_{3}$ measured at various applied magnetic fields. Figure 4(b) indicates that the XMCD peak intensity remains high down to an applied field of 0.1 T indicating that ferromagnetism exists in this sample.  The XMCD line shape is independent of the magnetic field. The slope of the the XMCD intensity $vs H$ curve indicates the paramagnetic part of the Fe magnetic moment.

\begin{figure}[htbp]
\begin{center}
\includegraphics[width=08cm]{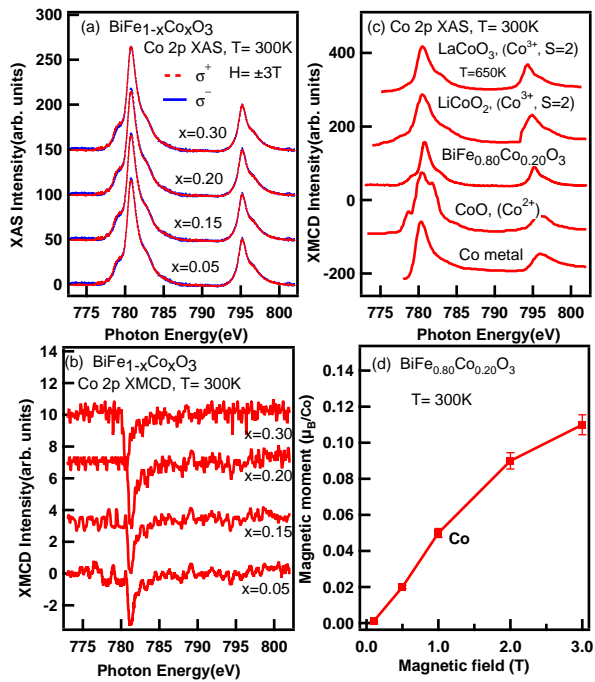}
\caption{(Color online) Co 2$p$ XAS and XMCD spectra of the BiFe$_{1-x}$Co$_x$O$_3$ ($x$ = 0 to 0.30) thin films (a) XAS spectra in magnetic field $\pm$ 3T at 300 K. (b) XMCD spectra. (c) XAS spectra of the BiFe$_{1-x}$Co$_x$O$_3$ thin films compared with the XMCD spectra of LaCoO$_3$ \cite{36}, LiCoO$_2$, CoO and Co metal \cite{35}. (d) Magnetic moment of Co in BiFe$_{0.8}$Co$_{0.2}$O$_3$ obtained from the XMCD intensities using sum rules as a function of magnetic field.}
\end{center}
\end{figure}

\ \ \ Figure 5(a) shows the Co 2$p$ XAS spectra of BFCO.  The corresponding XMCD data are shown in Fig. 5(b). Each spectrum consists of the $L_3$ (2$p_{3/2}$) edge at $\sim$ 781 eV and the $L_2$ (2$p_{1/2}$) edge at $\sim$ 795 eV. From comparison with the Co XAS spectra of other Co compounds shown in Fig. 5(c), it is clearly seen that the Co 2$p$ XAS spectrum of BFCO thin films is similar to that of LaCoO$_3$ at high temperatures ($\sim$ 650 K)high-spin Co$^{3+}$ and that of LiCoO$_2$ (high-spin Co$^{3+}$) but is different from that of CoO (high-spin Co$^{2+}$) \cite{35}. Thus we conclude that the Co ions in the BFCO films are mainly in the trivalent high-spin states \cite{36}. The spectral line shape of Co 2$p$ XAS did not change with $x$, indicating that the valence and spin states of the Co ions do not change with Co content. Similarly, as shown in Fig. 5(b), the XMCD peak intensity was also nearly independent of Co content. Figure 5(d) shows the magnetic moment of Co in BiFe$_{0.8}$Co$_{0.2}$O$_3$ obtained from the XMCD intensity as a function of magnetic field. The figure indicates that the Co magnetic moment is smaller and approaches zero as the magnetic field goes to zero. We consider that the Co$^{3+}$ ions are in the antiferromagnetic state, and attribute this weak XMCD signal to spin canting induced by the applied magnetic field. The magnetic moment of the Fe and Co ions at H = 3 T estimated as a function of Co content using the XMCD spin sum rule \cite{37} are shown in Fig. 3(b).


\ \ \ It has been revealed that the $M-H$ curves of the BFCO films were clearly saturated at H = 1T (not shown here). The magnetization for $x$ = 0.15 and 0.20 is enhanced and a finite coercive field (Hc) was observed however, the magnetization for $x >$ 0.20 decreases because of decreased $\gamma$-Fe$_2$O$_3$. For BFO, a strong coupling  between the ferroelectric 71$^\circ$ and 109$^\circ$ domains and the antiferromagnetic domains has been reported, and the anitiferromagnetic domains can be reversed by ferroelectric switching at RT \cite{38, 39}. In analogy with the BFO result, the BFCO films may potentially exhibit the magnetoelectric (ME) effect with a change in the macroscopic magnetization because only rhombohedral domains exist for Co concentration below $x$ =0.20.

\section{Conclusion}

\ \ \ We have performed XAS and the XMCD measurements of BFCO thin films, which exhibit ferromagnetism at 300 K. From analysis of the line shapes and intensity of Fe 2$p$ XMCD, we conclude that the ferromagnetism originates from both the $\gamma$-Fe$_2$O$_3$ impurity and the Fe$^{3+}$ ions in the BFCO lattice, and that the latter contribution becomes dominant for high Co content $x >$ 0.2, i.e., in the tetragonal-like monoclinic crystal structure. The Co ions are in the trivalent high-spin states and are most likely antiferromagnetic with weak magnetic moment due to spin canting induced by the applied magnetic field. The present results suggest that the Co substitution enhance the ferromagnetic behavior of Fe through the increased canting angle of the Fe$^{3+}$  spins in the rhombohedral phase with increasing $x$.

\begin{acknowledgments}

\ \ \ The experiment at SPring-8 was performed at the JAEA beamline BL23-SU (Proposal No. 2010A3826). This work was supported by a Grant-in-Aid for Scientific Research (S22224005) from JSPS, Japan.
\end{acknowledgments}

\appendix
\section{}

\setcounter{figure}{0}

\renewcommand{\thefigure}{A.\arabic{figure}}

\begin{figure}
\begin{center}
\includegraphics[width=06.5cm]{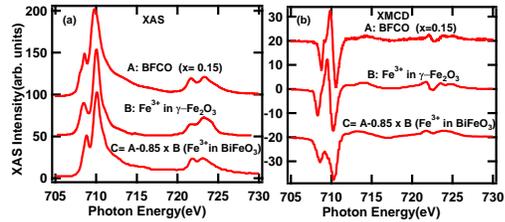}
\caption{(Color online) Decomposition of the Fe 2$p$ XAS (a) and XMCD (b) spectra of the BiFe$_{0.85}$Co$_{.15}$O$_3$ film (A) into $\gamma$-Fe$_2$O$_3$-like component (B), and the Fe$^{3+}$ in the BCFO lattice. The bottom spectrum (C) has been obtained by subtracting an appropriate amount of the $\gamma$-Fe$_2$O$_3$ spectrum from the BiFe$_{0.85}$Co$_{.15}$O$_3$ spectrum (C = A $-$0.85 $\times$ B) so that the result most resembles the Fe 2$p$ XAS and XMCD spectra of GaFeO$_3$ (Ref.\cite{33}).}
\end{center}
\end{figure}

In the Fe 2$p$ XAS and XMCD spectra, signals from Fe$^{3+}$ ($T_d$) and those from Fe$^{3+}$($O_h$) show different composition dependencies as shown in Fig. 2(d). We interpret this to the overlapping signals of the $\gamma$-Fe$_2$O$_3$-like impurity phase and Fe$^{3+}$ ($O_h$) in the BFO lattice. In order to demonstrate this, we have decomposed the measured spectra into the spectra of $\gamma$-Fe$_2$O$_3$ and Fe$^{3+}$ ($O_h$) of BFCO by subtracting the Fe 2$p$ XAS and XMCD spectra of $\gamma$-Fe$_2$O$_3$ as shown in Fig. A1. The subtraction was made so that the deduced spectrum becomes similar to the Fe 2$p$ XAS and XMCD of GaFeO$_3$ \cite{33}. Figure A2 shows the XMCD spectra from $x$ = 0 to 0.30 and their decomposition into the spectra of Fe$^{3+}$ in the BFO lattice and  those of $\gamma$-Fe$_2$O$_3$- like phase.

\begin{figure}
\begin{center}
\includegraphics[width=08cm]{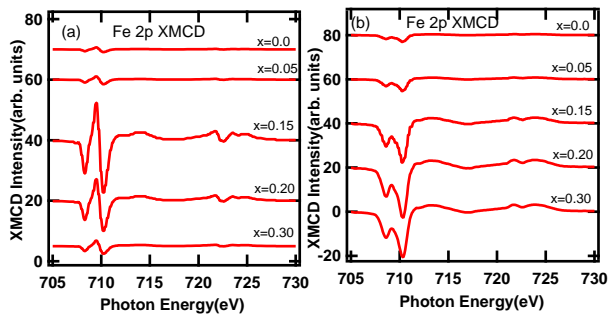}
\caption{(Color online) Decomposition of Fe 2$p$ XMCD spectra of BiFe$_{1-x}$Co$_x$O$_3$ thin films (a) into the $\gamma$-Fe$_2$O$_3$ -like component (b) and the Fe$^{3+}$ component in the BFCO lattice.}
\end{center}
\end{figure}

\end{document}